\documentclass{elsart}
\usepackage{epsfig}
\newcommand{\bm}[1]{ \mbox{\boldmath $#1$}  }

\begin{document}

\begin{frontmatter}

\title{Neutron-$^3$H potentials and the $^5$H-properties}

\author{R. de Diego \and E. Garrido} 
\address{ Instituto de Estructura de la Materia, CSIC, 
Serrano 123, E-28006 Madrid, Spain }

\author{D.V. Fedorov \and A.S.~Jensen}
\address{ Department of Physics and Astronomy,
        University of Aarhus, DK-8000 Aarhus C, Denmark }

\date{\today}

\maketitle

\begin{abstract}
The continuum resonance spectrum of $^5$H ($^3$H+$n$+$n$) is
investigated by use of the complex scaled hyperspherical adiabatic
expansion method. The crucial $^3$H-neutron potential is obtained by
switching off the Coulomb part from successful fits to $^3$He-proton
experimental data.  These two-body potentials must be expressed
exclusively by operators conserving the nucleon-core mean field
angular momentum quantum numbers.  The energies $E_R$ and widths
$\Gamma_R$ of the $1/2^+$ ground-state resonance and the lowest two
excited $5/2^+$ and $3/2^+$-resonances are found to be $(1.6,1.5)$
MeV, $(2.8,2.5)$ MeV and $(3.2,3.9)$ MeV, respectively.  These results
agree with most of the experimental data.  The energy distributions of
the fragments after decay of the resonances are predicted.
\end{abstract}

\end{frontmatter}

\par\leavevmode\hbox {\it PACS:\ } 21.45.+v, 31.15.Ja, 25.70.Ef

\maketitle

\section{Introduction}

Recent advances in experimental techniques have opened the door to
investigate superheavy hydrogen isotopes, $^5$H and $^7$H.  None of
them, nor $^4$H, have bound states, while $^3$H is well bound with the
neutron separation energy of 6.26~MeV. At the neutron dripline, where
one neutron becomes unbound, the structure has been successfully
described as an ordinary nuclear core surrounded by weakly bound or
unbound neutrons \cite{jen04}.  It is therefore natural to describe
the $^A$H nuclei ($A>3$) as a bound triton-core, denoted $t$ or $^3$H,
surrounded by $A-3$ neutrons.

In the present work we focus on $^5$H ($t$+$n$+$n$), for which the
experimental data are controversial. A summary of the different
experimental data and techniques can be found in \cite{gri04}.
In \cite{kor01} the $^5$H ground
state is found to be a 1/2$^+$ state with energy $E_R=1.7\pm0.3$ MeV
and width $\Gamma_R=1.9\pm0.4$ MeV. Similar results were found in
\cite{gol05}, where the authors quote an energy of 1.8 MeV and a
width of 1.3 MeV, and in \cite{ter05}, where the 1/2$^+$ ground state
of $^5$H is located at 2 MeV with a width of 2.5 MeV. In
\cite{mei03a,mei03b} a broad structure is observed in the $^5$H energy
distribution after proton knockout from $^6$He, that was interpreted
as a $1/2^+$ resonance at around 3 MeV. No evidence for a narrow
resonance in the $t+n+n$ system was obtained in contrast to
\cite{sid03}, where two rather narrow resonances were reported at
$1.8\pm0.1$~MeV and $2.7\pm0.1$~MeV with widths less than $0.5$~MeV.
In \cite{gur05} even a higher energy of $5.5\pm0.2$ MeV and a larger
width of $5.4\pm0.6$ MeV are given for the $^5$H ground state, which
is consistent with the results presented in \cite{ale95}.

The data obtained for the excited 3/2$^+$ and 5/2$^+$ states tend to
agree that these resonances are broad structures, almost degenerate,
with energies varying between 2.5 MeV \cite{gol05,ter05} and more than
10 MeV \cite{gur05}.  Different theoretical calculations concerning
$^5$H are available \cite{fil99,shu00,des01,tim02,ara03}. In general
they agree in placing the 1/2$^+$ ground state between 2 and 3 MeV
except \cite{fil99} where 6 MeV is reported.  The widths of these
resonances are only rough estimates except for the more complete
computation in \cite{ara03}.

The first three-body calculation of $^5$H \cite{shu00} used the
hyperspheric harmonic expansion method with a neutron-triton
interaction obtained from fits to the scarce amount of experimental
phase shifts \cite{shu96}.  The interaction is not consistent with the
available $^3$He-proton phase shifts.  The resonances are obtained
from an analysis of the three-body phase shifts and the bumps observed
in the computed missing mass spectra.  The authors find a ground state
energy at around 2.5-3.0 MeV.  However extraction of the resonance
properties from the missing mass spectra is not unambiguous, since it
requires a full description of the process used to populate the
$^5$H-states (initial state, reaction mechanism, final state
interactions).  This problem is discussed in detail in
\cite{gri04b}, which shows how different descriptions of the reaction
process can provide different properties of the $^5$H ground state.

The origin of these uncertainties is often related to a mismatch
between experimental and theoretical definitions of resonances.  The
experimental analyses of resonance energies and widths are most often
consistent with the definition of resonances as generalized
eigenstates of a given system, i.e. when they are defined as poles of
the $S$-matrix in the fourth quadrant of the complex energy
plane. With this assumption the reaction mechanism used to populate the
resonances becomes unimportant. The only essential ingredients in
the calculation of three-body resonances are then the different
two-body interactions.

The main theoretical problem is that the resonance wave functions
diverge asymptotically, which makes them difficult to disentangle from
ordinary non-resonant continuum wave functions. Different methods are
available to overcome this problem. For instance, the analytic
continuation of the coupling constant \cite{tan97} is used in
\cite{des01} to investigate $^5$H, and the ground state is found at
around 3 MeV with a width varying between 1 and 4 MeV.

An efficient method to obtain the $S$-matrix poles is the complex
scaling method \cite{ho83,moi98}, where all spatial coordinates are
rotated into the complex plane by a judicially chosen angle
$\theta$. In this way, provided that $\theta$ is larger than the
argument of the resonance, the wave function of the resonance falls
off exponentially exactly as for a bound state. In \cite{ara03} the
complex scaling method together with the microscopic three-cluster
model is used to investigate the $^5$H nucleus. An effective
nucleon-nucleon interaction is used to construct the neutron-triton
potential.  The energy and width of the 1/2$^+$ ground state are then
found to be 1.59 MeV and 2.48 MeV, respectively, while the excited
3/2$^+$ and 5/2$^+$ states are found almost degenerate at 3.0 MeV, and
with widths above 4 MeV.

In the last years it has been shown that the hyperspherical adiabatic
expansion method \cite{nie01} is an efficient tool to obtain
three-body bound states and resonances when combined with the complex
scaling method. The method has been applied successfully to
investigate systems like $^6$He, $^6$Be, $^6$Li, $^{11}$Li, $^{12}$C,
and $^{17}$Ne, \cite{gar02,fed03,gar04,gar05}, where the agreement
with the available experimental data is found to be remarkably good.
We therefore believe that a similar investigation of the properties of
$^5$H can help to clarify the existing uncertainties.

The paper is organized as follows: In section \ref{sec3} we introduce
the method and discuss in detail the two-body neutron-triton
interaction, that is the most crucial ingredient in the
calculation. In section \ref{sec4h} the properties of $^4$H are discussed.
The three-body results are shown in section
\ref{sec4}.  The energy distributions of the particles after decay of
the three-body resonances are given in section \ref{sec6}. We
then finish in section \ref{sec7} with the summary and the
conclusions.

\section{Crucial ingredients}
\label{sec3}

The adiabatic expansion method with hyperspherical coordinates and the
Faddeev decomposition is described for three-body systems in
\cite{nie01}.   The extension to compute resonances with complex rotation 
for these systems is described in \cite{fed03}.  We shall here specify
the details necessary to understand the subsequent discussion. First
we sketch the method and then we discuss in detail the decisive
neutron-triton interaction.

\subsection{Notation and parameters}

We define the hyperradius $\rho$ by
\begin{eqnarray}  
   m (m_1 + m_2 +  m_3) \rho^{2} =  \sum^{3}_{i<j} m_i m_j
 \left(\bm{r}_{i}-\bm{r}_j\right)^{2}  \label{e120} \;,
\end{eqnarray}
where $m_i$ and $\bm{r}_{i}$ are the mass and coordinate of particle
number $i$.  The mass $m$ is arbitrary and here chosen as the nucleon
mass.  All other relative coordinates are dimensionless angles
collectively denoted by $\Omega$.  The hyperradius is rotated by an
angle $\theta$ into the complex plane by multiplication with
$\exp(i\theta)$.

The total wave function $\Psi^{(JM)}$ with total angular momentum,
$J$, and projection, $M$, is expanded on the complete set
$\Phi^{(JM)}_n$ of adiabatic angular wave functions obtained for a
fixed value $\rho$
\cite{nie01}, i.e.
\begin{equation} \label{e125}
\Psi^{(JM)} = \frac{1}{\rho^{5/2}}\sum_n 
 f_n (\rho) \Phi_{n}^{(JM)} (\rho,\Omega) 
\end{equation}
\begin{equation}
 \Phi_{n}^{(JM)} (\rho,\Omega) =
 \sum_{i=1}^{3} \phi_{n}^{(i)} (\rho,\Omega_i) 
\end{equation}
where $\phi_{n}^{(i)}$ is the Faddeev component related to the Jacobi
system labeled by $i$.  The expansion coefficients are the radial
wave functions, $f_n(\rho)$, obeying a coupled set of radial equations
obtained by projecting the complex scaled Faddeev equations on the
adiabatic angular wave functions.  They are exponentially decaying for
resonances when the rotation angle $\theta$ of the hyperradius is
larger than that corresponding to the three-body resonance.  Then both
real and imaginary parts of the resonance energy, $E_0=E_R-iE_I$, ,
are determined by the boundary condition of $f_n$, i.e.
\begin{eqnarray}
 && f_{n}(\kappa\rho e^{i\theta})  \rightarrow
\sqrt{\rho} H_{K+2}^{(1)}(|\kappa| \rho e^{i(\theta-\theta_R)})
   \label{eq7} \\  
 &&\rightarrow
     \exp\Big({-|\kappa|\rho \sin{(\theta-\theta_R)}} 
 + i\left(|\kappa|\rho
\cos{(\theta-\theta_R)}-K\pi/2+3\pi/4 \right)\Big), \nonumber
\end{eqnarray}
where $\theta_R$ is the argument of the complex momentum,
$\kappa=\sqrt {2mE_0/\hbar^2}= |\kappa|\exp(-i\theta_R)$, and
$H_{K+2}^{(1)}$ is the Hankel function of the first kind.

The two-body interaction is the all-decisive input. In general, the
low-energy scattering properties of all pairs of particles should be
reproduced.  A smaller amount of data like scattering length and
effective range, or perhaps the low-lying two-body resonance energies
and their widths may also be sufficient.

The neutron-neutron interaction is well established. We use the
nucleon-nucleon potential given in \cite{gar99} which reproduces the
experimental $s$- and $p$-wave scattering lengths and effective
ranges. It contains central, spin-orbit ($\bm{\ell}\cdot\bm{s}$),
tensor ($S_{12}$) and spin-spin ($\bm{s}_1\cdot\bm{s}_2$)
interactions.  As indicated in \cite{zhu93} and supported by previous
calculations with the hyperspheric adiabatic expansion method, the
particular radial shape is not essential in descriptions of weakly
bound systems like $^6$He or $^{11}$Li as long as the low energy
scattering parameters are correct.

The interactions include an effective three-body force, $V_{3b}$,
which is necessary for fine-tuning to experimental energies, since
three-body states otherwise typically are underbound, and the precise
energy is crucial for the size and width.  Unfortunately, except for
the short-range character, the detailed properties of such interaction
are unknown and therefore create a source of uncertainties.  However,
precisely due to the short-range character of this interaction, we
expect that the more spatially extended the system is, the smaller is
the effect of the three-body potential.  We choose a Gaussian
dependence on hyperradius, $V_{3b}= S_{3b} \exp(-\rho^2/b_{3b}^2)$,
where the strength and the range are determined to fine-tune the
results if necessary. When $V_{3b}$ is diagonal and the same for all
adiabatic potentials, the partial wave structure of the three-body
states is maintained, but the energy can be adjusted.  We then expect
that the effects of $V_{3b}$ on the three-body resonances can be only
marginal when we maintain the same three-body energy.

\subsection{Neutron-triton potential}

The neutron-triton interaction is an important source of
uncertainties, since the amount of experimental data concerning $^4$H
are rather scarce.

In \cite{til92} four resonances in $^4$H are reported to have energies
and widths $(E_R,\Gamma_R)$ of $(3.19,5.42)$ MeV ($2^-$), $(3.50,
6.73)$ MeV ($1^-$), $(5.27,8.92)$ MeV ($0^-$), and $(6.02,12.99)$ MeV
($1^-$).  These energies are not well established, and more recent
experimental analyses suggest that they could be smaller, placing the
ground state of $^4$H at around 2 MeV \cite{gur05}.   Data in both
\cite{til92} and \cite{gur05} are obtained by a Breit-Wigner fit to
experimental cross sections or missing mass spectra.

The neutron-triton interaction should reproduce the experimental
observables either directly by computation or indirectly by comparing
the same derived quantities. Matching computed $S$-matrix poles to the
experimental Breit-Wigner parameters from \cite{til92} or \cite{gur05}
is then not an appropriate procedure.  A better method is to construct
the neutron-triton interaction by reproducing the experimental phase
shifts.  Unfortunately only very few experimental phase shifts are
available, and furthermore their error bars are large \cite{tom66}.
On the other hand much more is known about the mirror system,
$^3$He-proton, where abundant and reliable data are available.
Invoking charge symmetry of the nuclear forces we can then construct a
$^3$He-proton potential and subsequently find the corresponding
neutron-triton potential by switching off the Coulomb interaction.
This procedure was followed for instance in \cite{des01}.

The experimental $^3$He-proton phase shifts \cite{tom65,bel85} are
usually quoted specifying their quantum numbers $\{\ell, s, j\}$,
where $\bm{\ell}$ is the $^3$He-proton relative orbital angular
momentum, $\bm{s}=\bm{s}_N+\bm{s}_c$ is the sum of the spins of the
proton, $\bm{s}_N$, and $^3$He, $\bm{s}_c$, and
$\bm{j}=\bm{\ell}+\bm{s}$ is the total two-body angular momentum.  It
is then tempting to consider the two particles on equal footing like
for the nucleon-nucleon interaction. This leads to a $^3$He-proton
interaction of the form:
\begin{equation}
 V_{Nc}(r)=V_c(r)+V_{ss}(r) \bm{s}_N\cdot \bm{s}_c + 
       V_{so}(r) \bm{\ell}\cdot(\bm{s}_N+\bm{s}_c)
\label{eq9}
\end{equation}
for which $\{\ell, s, j\}$ are conserved quantum numbers. This kind of
potential has for instance been used in \cite{shu00} for the
neutron-triton potential.

The potential in Eq.(\ref{eq9}) has the problem that $s$ and $j$ are
not the mean-field quantum numbers of the nucleons within the
$^3$He-core, where every nucleon moves in an orbit characterized by
the relative nucleon-core orbital angular momentum $\ell$ and the
total angular momentum of that nucleon $j_N=\ell \pm 1/2$.  For the
neutron-core system $\bm{j}_N$ couples to the core-spin,
$\bm{s}_c$, to give the total two-body angular momentum $\bm{j}$.  The
problem arises because the nucleon angular momentum $j_N$ is not
conserved for the potential in Eq.(\ref{eq9}). Therefore the
mean-field spin-orbit partners of the nucleons within the core with
$j_N=\ell \pm 1/2$ are inevitably mixed.  The motion of the valence
nucleon(s) outside the core is then in conflict with the (approximate)
mean-field motion of the identical nucleons within the core.  This
problem is discussed in detail in \cite{gar03}. These inconsistencies
are enhanced if one and only one of the mixed partners is occupied in
the core by valence nucleons. An example of such Pauli forbidden
states is the $p_{3/2}$ orbit in $^{10}$Li and $^{11}$Li \cite{gar03}.

This particular violation of the Pauli principle is not present in
systems like $^3$H and $^3$He, but certainly the nucleons outside the
core should preferably occupy the $p_{3/2}$ states instead of the
$p_{1/2}$ levels (or the $d_{5/2}$ instead of the $d_{3/2}$).  This is
not possible with the potential in Eq.(\ref{eq9}), but we can achieve
full consistency with the mean-field description by replacing the
potential in Eq.(\ref{eq9}) by \cite{gar99}:
\begin{equation}
 V_{Nc}(r)=V_c(r)+V_{js}(r) (\bm{\ell}+\bm{s}_N)\cdot \bm{s}_c 
 + V_{so}(r) \bm{\ell}\cdot\bm{s}_N \; ,
\label{eq10}
\end{equation}
which has $\{\ell, j_N, j\}$ as conserved quantum numbers, where
$\bm{j}_N=\bm{\ell}+\bm{s}_N$ and $\bm{j}=\bm{j}_N+\bm{s}_c$.  This
choice of the $^3$He-proton interaction in Eq.(\ref{eq10}) is also
supported by the experimental data.  In \cite{tom65}, and especially
in the more recent data \cite{bel85}, a singlet-triplet mixing in the
$^3$He-proton $p$-states was found experimentally.  Such mixing can
be achieved with the potential in Eq.(\ref{eq10}), but not with
Eq.(\ref{eq9}).  In fact, in \cite{bel85} it is proposed that the mixing
can be explained by the spin-orbit operator, $\bm{\ell} \cdot
\bm{s}_N$, precisely as suggested in \cite{gar03} and expressed in
Eq.(\ref{eq10}).

For $\ell\neq0$ this spin-orbit operator in Eq.(\ref{eq10}) gives rise
to two sets of degenerate states $\{\ell_{j_N=\ell-1/2}^{(j=\ell-1)},
\ell_{j_N=\ell-1/2}^{(j=\ell)}\}$ and
$\{\ell_{j_N=\ell+1/2}^{(j=\ell)},
\ell_{j_N=\ell+1/2}^{(j=\ell+1)}\}$. This degeneracy is broken by the
term, $\bm{j}_N\cdot \bm{s}_c$, of the potential. For
instance, for $p$-waves in $^4$H the spin-spin term of the potential
breaks the degeneracy between the $\{p_{3/2}^{(j=1)},
p_{3/2}^{(j=2)}\}$ and the $\{p_{1/2}^{(j=0)}, p_{1/2}^{(j=1)}\}$
states.  Therefore, one should find in $^4$H two relatively
close-lying 1$^-$ and 2$^-$ states, separated by a relatively large
energy gap from another couple of close-lying 0$^-$ and 1$^-$
states. This structure is precisely the one observed in \cite{til92}
for the resonances in $^4$H.  This also supports potential
(\ref{eq10}) over potential (\ref{eq9}) where the latter instead would
produce two sets of $p$-states in $^4$H with rather similar energy
separation \cite{gar03}.

Therefore we have constructed a nuclear $^3$He-proton interaction of
the form in Eq.(\ref{eq10}). We have taken the central, $V_c(r)$,
$js$, $V_{js}(r)$, and spin-orbit, $V_{so}(r)$, radial form
factors as a sum of two Gaussians. The strengths and ranges of each
Gaussian have been adjusted such that after adding the Coulomb
potential the experimental phase shifts for $s$, $p$, and $d$-waves
from \cite{bel85} are reproduced.  As shown in \cite{bel85,dri70}
the assignment between the $\delta_{\ell,s}^j$ phase shifts given
experimentally in terms of the quantum numbers associated to the
potential (\ref{eq9}) and the ones associated to potential
(\ref{eq10}) are the following: $p_{3/2}^{(j=1)}\rightarrow
\delta_{1,1}^1$, $p_{3/2}^{(j=2)}\rightarrow \delta_{1,1}^2$,
$p_{1/2}^{(j=1)}\rightarrow \delta_{1,0}^1$, and
$p_{1/2}^{(j=0)}\rightarrow \delta_{1,1}^0$ for $p$-waves, and
$d_{5/2}^{(j=3)}\rightarrow \delta_{2,1}^3$,
$d_{5/2}^{(j=2)}\rightarrow \delta_{2,1}^2$,
$d_{3/2}^{(j=2)}\rightarrow \delta_{2,0}^2$, and
$d_{3/2}^{(j=1)}\rightarrow \delta_{2,1}^1$ for $d$-waves. For
$s$-waves both potentials (\ref{eq9}) and (\ref{eq10}) are obviously
equivalent.

\begin{table}
\begin{center}
\caption{Strengths $S_i$ and ranges $b_i$ of the two Gaussians 
for the central ($V_c^{(\ell)}$), $js$-term ($V_{js}^{(\ell)}$), and
spin-orbit ($V_{so}^{(\ell)}$) potentials in Eq. (\ref{eq10}).}
\vspace*{0.5cm}
\begin{tabular}{|c|ccc|ccc|ccc|}
\hline
   &
\multicolumn{3}{c|}{$V_c^{(\ell)}$ } &
\multicolumn{3}{c|}{$V_{js}^{(\ell)}$ } &
\multicolumn{3}{c|}{$V_{so}^{(\ell)}$ } \\
\hline
$\ell$ & $0$ & $1$ & $2$ & $0$ & $1$ & $2$ & $0$ & $1$ & $2$ \\
\hline
$S_1$ (MeV) & 20.17 & -400.01 & 10.04 & -1.45 & -237.77 & 2.80 & -- & -135.98 & -2.15 \\
$b_1$ (fm) & 3.59 & 2.62   & 3.62 & 7.39 & 1.44   & 3.38 & -- & 2.09   & 3.68 \\
$S_2$ (MeV) & --   & 400.81 & --   & --   & 108.87 & --   & -- & 197.56 & -- \\
$b_2$ (fm) & --   & 2.55   & --   & --   & 1.70   & --   & -- & 1.82   & -- \\
\hline
\end{tabular}
\label{tab1}
\end{center}
\end{table}

\begin{figure}
\vspace*{-0.8cm}
\epsfig{file=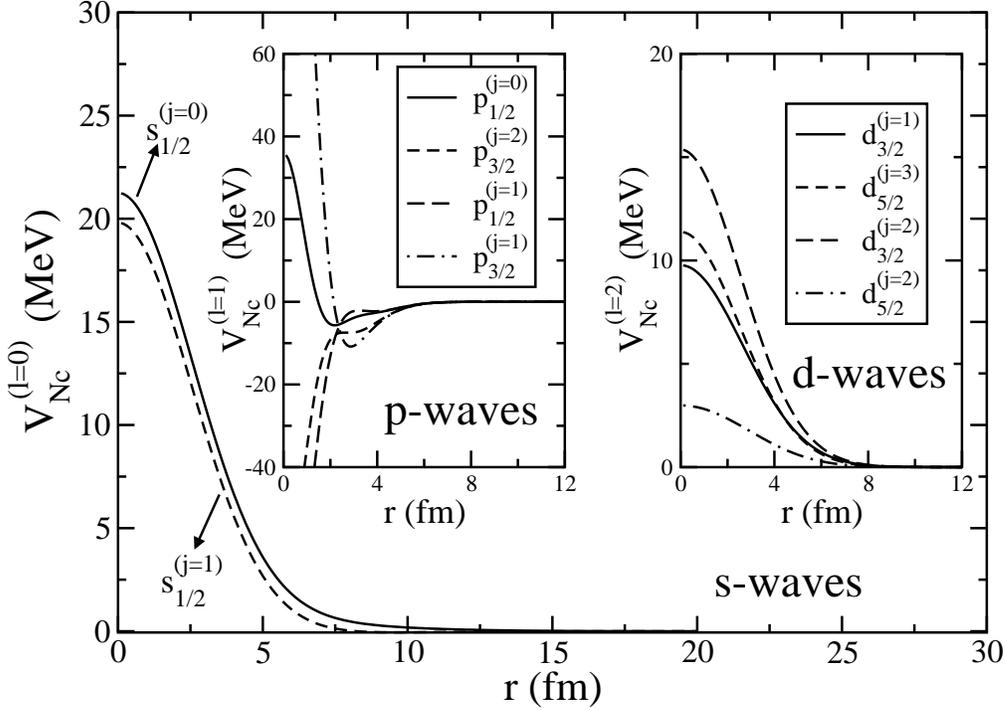, width=11cm, angle=-90}
\vspace*{-0.5cm}
\caption{Neutron-$^3$H potentials for the different $\ell_{j_N}^{(j)}$ 
waves corresponding to the parameters in table~\ref{tab1}. The
external panel shows the two $s$-wave potentials, and the internal panels
show the four $p$-wave (left), and the four $d$-wave (right)
neutron-$^3$H potentials, respectively.}
\label{fig1}
\end{figure}

Using a gradient minimization method we have found the
strengths and ranges of the Gaussian potentials given in table
\ref{tab1}. In Fig.\ref{fig1} we plot the corresponding $^3$H-neutron
potentials for the different $\ell_{j_N}^{(j)}$ states.  The $s$-wave
potentials are repulsive, which is consistent with population by two
protons or neutrons of the two $s_{1/2}$ states in $^3$He or $^3$H,
respectively.  As a consequence the third proton in $^4$Li, or the
third neutron in $^4$H, can not populate these $s$-states due to the
Pauli principle.  The potentials for the $d$-states are also
repulsive, pushing up the energies of the $d_{5/2}$ and $d_{3/2}$
states. This is also expected but for a different reason, i.e. the
$d$-states should be higher in energy than the states in the $p$
shell, and therefore with a smaller probability of being populated by
the valence nucleon.

\begin{figure}
\vspace*{-1.5cm}
\epsfig{file=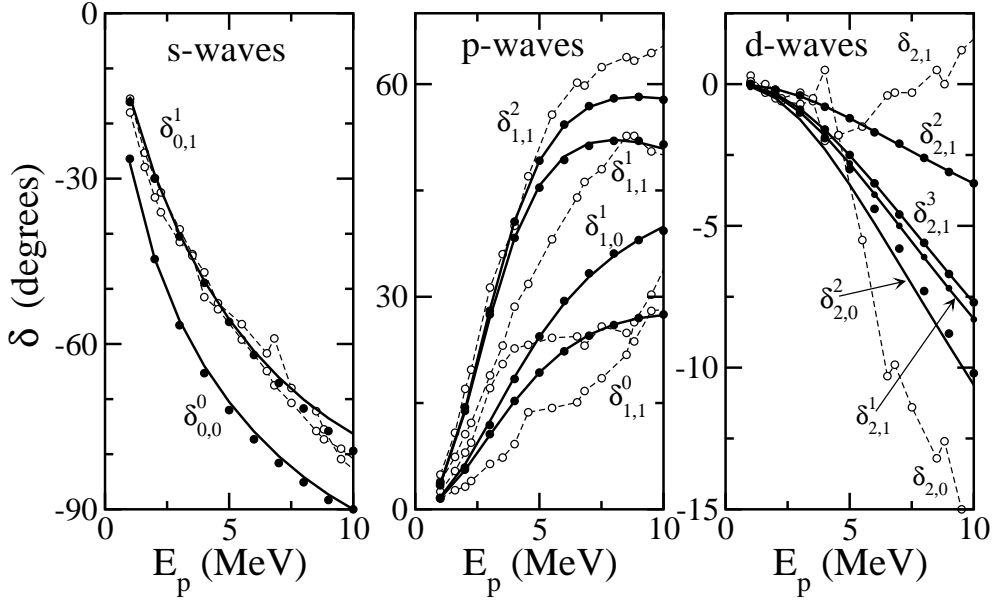, width=11cm, angle=-90}
\vspace*{-1.5cm}
\caption{Computed phase shifts $\delta_{\ell,s}^j$ (solid lines) for 
the $s$-wave (left), $p$-wave (center), and $d$-wave (right)
$^3$He-proton potentials using the parameters given in table
\ref{tab1}. The solid circles show the experimental phase shifts in
\cite{bel85}, and the open circles (connected with dashed curves) are
the experimental data from \cite{tom65}. In the right part of the
figure the solid circles correspond to $\delta_{2,0}^2$ from
\cite{bel85} but with opposite sign.}
\label{fig2}
\end{figure}

In Fig.\ref{fig2} we show the computed phase shifts (solid lines) for
the different $s$, $p$ and $d$-wave $^3$He-proton potentials. The
agreement with the experimentally derived phase shifts in \cite{bel85}
(black circles) is very good. For comparison we also show (open
circles) the experimental phase shifts given in \cite{tom65}. To guide
the eye we have connected these data with dashed curves. These data
\cite{tom65} are not fully consistent with the ones given in
\cite{bel85}, although the global behaviour is similar. Also, in
\cite{tom65} the $d$-wave phases are assumed to be degenerate (denoted 
$\delta_{2,0}$ and $\delta_{2,1}$ in the right part of Fig.\ref{fig2})
with a resulting erratic dependence on the energy.

The present numerical fit for $d$-waves only used the experimental
$\delta_{2,1}^1$, $\delta_{2,1}^2$, and $\delta_{2,1}^3$ phase shifts,
since the numerical method was unable to find a potential matching
simultaneously all  four sets of $d$-wave shifts in \cite{bel85}.
As seen in the right part of Fig.\ref{fig2} the match of the three
sets of experimental phase shifts used for the numerical fit (black
circles) is excellent.  The lowest solid line shows the
$\delta_{2,0}^2$ phase shifts obtained with the potential parameters
fitting the other three sets. It is puzzling that these phase shifts
are in almost perfect agreement with those in \cite{bel85}, provided
the sign is reversed (black points).  In any case these computed
$\delta_{2,0}^2$ phase shifts exhibit the same global behaviour
(negative and decreasing with a similar rate) as those of \cite{tom65}
(open circles).

We have also tried to fit the experimental $^3$He-proton phase shifts
\cite{bel85} with the potential in Eq.(\ref{eq9}), even though the 
conserved quantum numbers then are inappropriate.  The numerical
procedure fails dramatically in determining the potential parameters
for the $p$ and $d$-waves in this case.  The same failure appears with
other analytical expressions for the radial form factors.  It is
certainly very difficult (if not impossible) to construct a potential
like Eq.(\ref{eq9}) reproducing the experimental $^3$He-proton phase
shifts. In total, this demonstrates that the experimental phase shifts
are inconsistent with Eq.(\ref{eq9}), but consistent with
Eq.(\ref{eq10}).

\section{$^4$H properties.}
\label{sec4h}

The structure of $^4$H is obtained from the neutron-$^3$H potentials given 
in table~\ref{tab1} and plotted in Fig.\ref{fig1}. Only the potentials 
in the $p_{3/2}$ and $p_{1/2}$ states exhibit attractive pockets with barriers.
These states are therefore the only ones which might support
two-body resonances. The central $p$-wave potential contains 
two Gaussians with slightly different ranges and almost
identical strengths of opposite sign, which is an attractive
surface central potential. We test the dependency on two choices of the Coulomb interaction
corresponding to Gaussian and point-like $^3$He-charge distributions. The
result in table \ref{tab1} for the point-like Coulomb potential is
almost identical to that of a Gaussian charge distribution. The fits
to the data are equally good. With artificially small errors of 0.1
degrees on all phase shift data points the $\chi^2$ per data is around 8.
This means that an error of 1 degree would give a $\chi^2$ per data point equal to about 0.08.

\begin{table}
\begin{center}
\caption{Resonance energies and widths $(E_R,\Gamma_R)$, in MeV, in $^4$H.
The 2$^{nd}$ and 3$^{rd}$ columns are the results obtained as poles of the $S$-matrix in this work and in 
\cite{ara03}. The 4$^{th}$ column gives the results from this work assuming that the resonances energies
correspond to a maximum of $\frac{d\delta}{d E}$ (and 
$\Gamma_R$=$2/\left(\frac{d\delta}{d E}\right)_{max}$). The last two columns are the experimental
values from \cite{til92} and \cite{gur05}, where an $R$-matrix fit or a Breit-Wigner fit to the peak of
the experimental cross sections is made. In \cite{gur05} the energies are given without spin-parity
assignment.}
\vspace*{0.5cm}
\begin{tabular}{|c|c|c|c|c|c|}
\hline
$J^\pi$ & This work &\cite{ara03}&  This work & \cite{til92}, exp. & \cite{gur05}, exp. \\
        & $S$-matrix poles& $S$-matrix poles&  $\left({d\delta}/{d E}\right)_{max}$ &
          $R$-matrix&  Breit-Wigner \\
\hline
   & $(E_R,\Gamma_R)$ & $(E_R,\Gamma_R)$ & $(E_R,\Gamma_R)$ & $(E_R,\Gamma_R)$ & $(E_R,\Gamma_R)$ \\
\hline
 2$^-$ & (1.22,3.34) & (1.52,4.11) &(1.80,6.21) & (3.19,5.42) & (1.6,$\sim$0.8) \\
 1$^-$ & (1.15,3.49) & (1.23,5.80) &(1.70,6.84) & (3.50,6.73) & (3.4,$\sim$0.8) \\
 0$^-$ & (0.77,6.72) & (1.19,6.17) &(1.92,19.1) & (5.27,8.92) & (6.0,$\sim$1.0) \\ 
 1$^-$ & (1.15,6.38) & (1.32,4.72) &(2.25,14.6) & (6.02,12.99) &  --- \\
\hline
\end{tabular}
\label{tab2}
\end{center}
\end{table}

The $p$-wave potentials give rise to four resonant states in $^4$H,
whose energies and widths are given in the second column of table~\ref{tab2}. 
These states have been obtained as poles of the $S$-matrix and agree
rather well with the ones found in \cite{ara03} (third column), 
where resonances are also computed as poles of the $S$-matrix, and where a
different procedure (resonating group method) is used to build the
neutron-triton interaction.

There is a clear disagreement between the theoretical energies and widths 
given in columns 2 and 3, and the experimental ones in column 5 \cite{til92}.
However, the experimental data \cite{til92} have been obtained
from a charge symmetric reflection of the $R$-matrix parameters for $^4$Li.
It is well known, as explicitly written in \cite{til92}, that the structure
given by the $S$-matrix poles is quite different, giving rise to $p$-wave resonances
occurring even in a different order. This fact is illustrated in 
\cite{cso97}, where large variations for the resonance parameters in $^5$He and $^5$Li 
are found depending on what procedure is used to extract them. In all cases
the resonance energies obtained as poles of the $S$-matrix are clearly
lower than the ones obtained using a conventional $R$-matrix analysis.
This is particularly important for broad resonances.

A recent Breit-Wigner fit \cite{gur05} of the experimental missing-mass spectra suggests
that the ground state energy of $^4$H is a factor of two lower than in \cite{til92} 
(last column in table~\ref{tab2}). The widths of all the states are also much smaller than the ones 
given in \cite{til92}. In \cite{gur05} only the energies and 
widths are given, without the spin and parity assignment.

\begin{figure}
\begin{center}
\epsfig{file=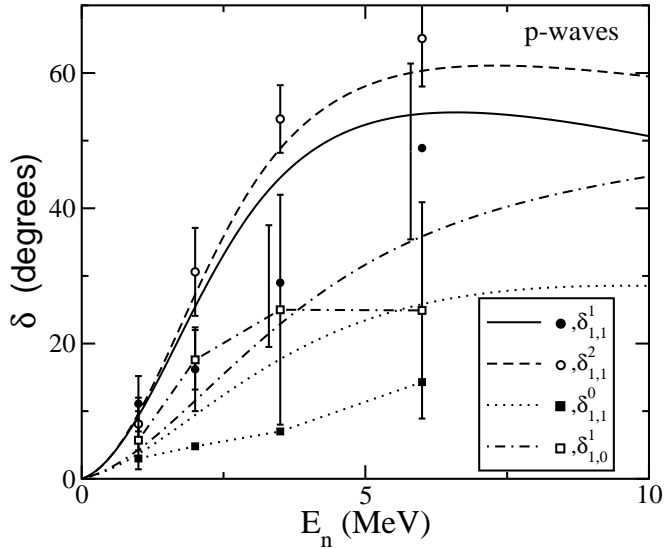, width=9cm, angle=-90}
\vspace*{-0.5cm}
\caption{Computed phase shifts $\delta_{\ell,s}^j$ for the $p$-wave neutron-$^3$H potential using 
the parameters given in table \ref{tab1}. The experimental data (solid circles for $\delta_{1,1}^1$,
open circles for $\delta_{1,1}^2$, solid squares for $\delta_{1,1}^0$, and open squares for 
$\delta_{1,0}^1$) are from \cite{tom66}.}
\end{center}
\label{fig3}
\end{figure}

It is quite obvious that a direct comparison of the resonance energies for $^4$H obtained
in this work as poles of the $S$-matrix with the ones obtained from the $R$-matrix or Breit-Wigner
analyses (given in the last two columns of table~\ref{tab2}) is not appropriate. This comparison 
can not be used to test the validity 
of the neutron-triton interaction. A more efficient way is to compare directly to the very
few available experimental neutron-triton phase-shifts \cite{tom66}. In Fig.\ref{fig3} we show the 
computed neutron-triton phase shifts ($\delta_{\ell,s}^j$) for the $p$-potentials in Fig.\ref{fig1}. 
The solid, dashed, dotted, and dot-dashed curves correspond to the $\delta_{1,1}^1$, $\delta_{1,1}^2$,
$\delta_{1,1}^0$, and $\delta_{1,0}^1$ phase shifts, respectively. The corresponding experimental
data \cite{tom66} are given by the solid circles, open circles, solid squares, and open squares, 
respectively. As seen in the figure, the error bars of the data are large. The $\delta_{1,1}^0$ data are only available without error bars, which presumably should be large. For $\delta_{1,1}^0$ and $\delta_{1,0}^1$ the experimental data have been 
connected with the same kind of curve as for the computed ones.

Using the parameters in table \ref{tab1} to compute (not fit) the very few and uncertain neutron-triton phase shifts
we arrive at an $\chi^2$ per data point equal to about 8, which is comparable to the $^3$He-proton fit. The neutron-triton phase shifts could easily
be precisely reproduced with potentials of the form in eq. (\ref{eq10}).
However, the few and inaccurate points do not make that worthwhile.
Instead we rely on the very precise fits of the $^3$He-proton phase
shifts and the charge independence of the strong interaction.

As seen in Fig.\ref{fig3}, the computed phase shifts agree reasonably well with the experiment. We can 
observe that neither the computed phase shifts nor the experimental ones cross the value 
$\pi/2$. The definition of resonances as the energy at which $\delta(E)$=$\pi/2$ is then not 
applicable here. This fact contradicts the results shown in \cite{des01}, where the neutron-triton
phase shifts do cross the value $\pi/2$. In fact, in \cite{des01} resonances are precisely
defined as the energies where $\delta(E)$=$\pi/2$. This seems to disagree with the
experimental data. The phase shifts in Fig.\ref{fig3} are consistent with the existence of broad
resonances defined as the energies where $d\delta/dE$ has a maximum, and  
$\Gamma_R$=$2/\left(\frac{d\delta}{d E}\right)_{max}$ \cite{car84}. Following this prescription
we obtain the energies and widths given in the fourth column of table \ref{tab2}. The different
states are now reordered, and the new energies and widths are significantly larger than the ones 
obtained as poles of the $S$-matrix. 

The $R$-matrix analysis of these experimental phase shifts \cite{tom66}
leads to a $2^-$ ground state energy for $^4$H of 3.4 MeV with a width of 5.5 MeV. 
Therefore, since our neutron-triton interaction agrees reasonably well with  
the experimental phase shifts \cite{tom66}, the resonance energies and widths obtained as 
poles of the $S$-matrix and given in the second column of table \ref{tab2}, are consistent with 
the $R$-matrix analysis which gives the  ground state energy slightly above 3 MeV.

The $s$-wave potential given in table \ref{tab1} gives 
rise to singlet and triplet scattering lengths equal to 5.32 fm and 3.01 fm, respectively,
in agreement with the experimental values of 4.98$\pm$0.29 fm and 3.13$\pm$0.11 fm, respectively
\cite{jon90}. Thus mirror symmetry seems to be rather well fulfilled.

\section{Results for $^5$H}
\label{sec4}

Once the two-body interactions are fixed, we investigate the
low-energy spectrum of $^5$H.  As discussed above only the attractive
$p$-waves can be responsible for low-lying states.  The resulting
angular momentum and parity, $J^{\pi}$, of these three-body states are
then expected to be 1/2$^+$, 3/2$^+$, and 5/2$^+$ states. The Pauli
principle prohibits larger $J$-values with positive parity.  Negative
parity states must involve one neutron-triton $p$-level and one of
either $s$ or $d$-character.  These three-body states are necessarily
then situated at higher energies if they appear at all.  For each
$J^{\pi}$, we first solve the angular part of the (complex rotated)
Faddeev equations, which provide the effective potentials in the
radial equations.  As a second step we solve the radial equations that
give the radial wave functions and the energies for each of the
states.

\subsection{Angular part.}
\label{subsec1}

To solve the angular part of the Faddeev equations, the angular
eigenfunctions $\phi_n^{(i)}$ in Eq.(\ref{e125}) are expanded in terms
of the complete basis $\left\{[Y_{\ell_x \ell_y}^{K L}\otimes
\chi_{s_x,s_y}]^J\right\}$, where the hyperspherical harmonics 
$Y_{\ell_x \ell_y}^{K L}$ are the free solutions with only the kinetic
energy operator and $\chi_{s_x,s_y}$ is the spin wave function.  For
each Jacobi set $\ell_x$ and $\ell_y$ are the orbital angular momenta
associated to the Jacobi coordinates $\bm{x}$ and $\bm{y}$, $s_x$ is
the coupling of the spins of the two particles connected by $\bm{x}$,
$L$ and $S$ are the total orbital angular momentum and spin,
respectively, and $J$ is the total angular momentum of the three-body
system.  In this expansion convergence is required at two different
levels. First, the basis containing all contributing partial wave
components $\{\ell_x,\ell_y,L,s_x,S\}$ must be included in the
expansion, and second, for each component, the maximum value,
$K_{max}$, of the hypermomentum $K$ must be large enough to ensure
convergence for all necessary distances.

\begin{table}[h]
\vspace*{0.5cm}
\begin{center}
\caption{The largest partial wave components included in the calculation for 
the 1/2$^+$ state of $^5$H. The left and right parts refer to the
Jacobi sets where $\bm{x}$ connects the two neutrons, or the core and
one of the neutrons, respectively. The sixth row gives the maximum
value of the hypermomentum $K$ used for each of these components.  The
last row gives the contribution of the component to the total norm of
the (complex rotated) wave function. Only those components
contributing more than 1\% are given. }
\vspace*{0.5cm}
\begin{tabular}{|lcccc|cccccc|}
\hline
$l_{x}$     & 0    & 1   & 1   & 2   & 0   & 0   & 1    & 1    & 1   & 1 \\
$l_{y}$     & 0    & 1   & 1   & 2   & 0   & 0   & 1    & 1    & 1   & 1 \\
$L$         & 0    & 1   & 1   & 0   & 0   & 0   & 0    & 0    & 1   & 1 \\
$s_{x}$     & 0    & 1   & 1   & 0   & 0   & 1   & 0    & 1    & 0   & 1 \\
$S$         & 0.5  & 0.5 & 1.5 & 0.5 & 0.5 & 0.5 & 0.5  & 0.5  & 0.5 & 1.5 \\
$K_{max}$   & 180  & 82  & 102 & 84  & 80  & 100 & 182  & 182  & 82  & 82 \\
$W$ ($^5$H) & 92.2 & 2.0 & 3.7 & 1.9 & 2.6 & 8.3 & 21.0 & 62.4 & 1.6 & 3.8 \\
\hline
\end{tabular}
\label{tab3}
\end{center}
\end{table}

\begin{table}[h]
\vspace*{0.5cm}
\begin{center}
\caption{The same as table \ref{tab3} for the 3/2$^+$ state of $^5$H}
\vspace*{0.5cm}
\begin{tabular}{|lcccc|cccccc|}
\hline
$l_{x}$     & 0    & 1   & 1    & 2    & 1   & 1   & 1    & 1    & 1    & 2 \\
$l_{y}$     & 2    & 1   & 1    & 0    & 1   & 1   & 1    & 1    & 1    & 0 \\
$L$         & 2    & 1   & 1    & 2    & 1   & 1   & 1    & 2    & 2    & 2 \\
$s_{x}$     & 0    & 1   & 1    & 0    & 0   & 1   & 1    & 0    & 1    & 1 \\
$S$         & 0.5  & 0.5 & 1.5  & 0.5  & 0.5 & 0.5 & 1.5  & 0.5  & 0.5  & 0.5 \\
$K_{max}$   & 182  & 82  & 122  & 122  & 82  & 82  & 122  & 122  & 182  & 82 \\
$W$ ($^5$H) & 43.8 & 5.0 & 20.1 & 30.6 & 3.8 & 1.3 & 19.6 & 18.2 & 52.6 & 4.0 \\
\hline
\end{tabular}
\label{tab4}
\end{center}
\end{table}

The 1/2$^+$, 3/2$^+$, and 5/2$^+$ states in $^5$H have been calculated
including all the possible components involving $s$, $p$, and
$d$-waves.  A $K_{max}$ value of at least 60 has been used, except for
the largest components where $K_{max}$ has been increased to guarantee
convergence of the effective potentials. In tables \ref{tab3},
\ref{tab4}, and \ref{tab5} we specify the quantum numbers, including
the $K_{max}$ values, of the largest components for each of the states
computed.

\begin{table}[h]
\vspace*{0.5cm}
\begin{center}
\caption{The same as table \ref{tab3} for the 5/2$^+$ state of $^5$H}
\vspace*{0.5cm}
\begin{tabular}{|lccc|ccccc|}
\hline
$l_{x}$     & 0    & 1    & 2    & 1    & 1    & 1    & 2   & 2 \\
$l_{y}$     & 2    & 1    & 0    & 1    & 1    & 1    & 0   & 0 \\
$L$         & 2    & 1    & 2    & 1    & 2    & 2    & 2   & 2 \\
$s_{x}$     & 0    & 1    & 0    & 1    & 0    & 1    & 0   & 1 \\
$S$         & 0.5  & 1.5  & 0.5  & 1.5  & 0.5  & 0.5  & 0.5 & 0.5 \\
$K_{max}$   & 182  & 82   & 122  & 82   & 122  & 182  & 82  & 82 \\
$W$ ($^5$H) & 55.1 & 13.1 & 31.4 & 12.7 & 19.9 & 60.1 & 1.8 & 5.2 \\
\hline
\end{tabular}
\label{tab5}
\end{center}
\end{table}

Solving now the angular eigenvalue equation we obtain the
eigenvalues, $\lambda_n(\rho)$, needed to construct the effective
potentials in the radial equations. In particular, it turns out that
it is sufficient to use a scaling angle of $\theta$=0.35 rads for the
$1/2^+$ and $3/2^+$ states, and 0.30 rads for the 5/2$^+$ state,
respectively.

\begin{figure}
\vspace*{-1.5cm}
\epsfig{file=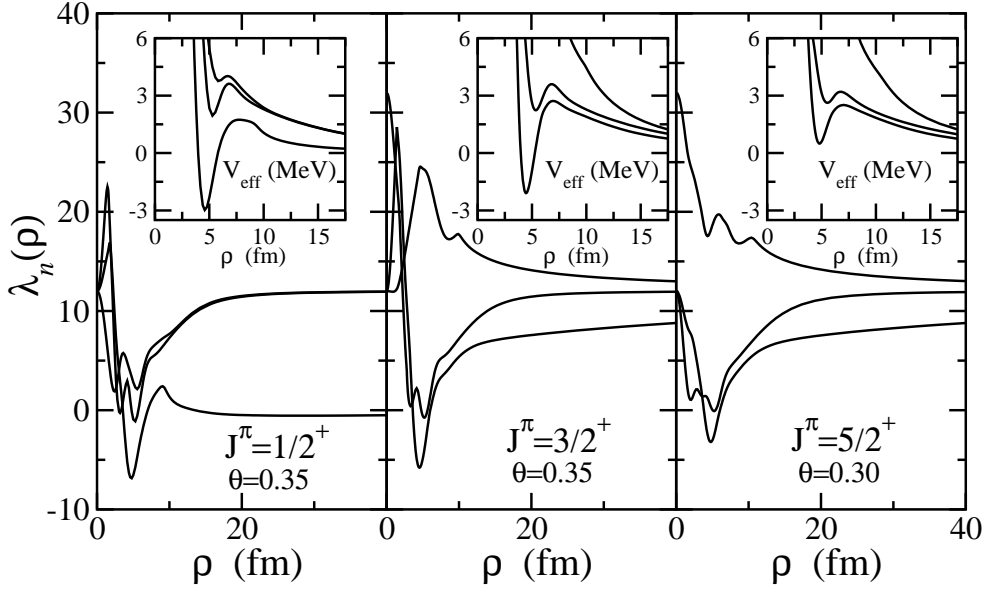,width=11cm, angle=-90}
\vspace*{-1.5cm}
\caption{Real parts of the three lowest angular eigenvalues $\lambda(\rho)$ 
(external plots) and the effective hyperradial potentials
$V_{eff}=\hbar^2(\lambda+15/4)/(2 m \rho^2)$ (insets) for the 1/2$^+$,
3/2$^+$ and 5/2$^+$ states of $^5$H as a function of $\rho$.}
\label{fig4}
\end{figure}

In Fig.\ref{fig4} we show the three most contributing eigenvalues
$\lambda(\rho)$ for each of the 1/2$^+$, 3/2$^+$ and 5/2$^+$ states in
$^5$H. For the sake of clarity, we only show the real parts, which for
short-range interactions must reproduce the hyperspherical spectrum
($K(K+4)$) at $\rho=0$ and $\rho=\infty$ \cite{fed03}.  The most
attractive pocket appears for the 1/2$^+$ state, which therefore is
expected to be the ground state, in agreement with the experimental
results. This is seen more clearly in the insets, which show the real
parts of the effective radial potentials,
$V_{eff}=(\lambda+15/4)/\rho^2$.  Together with the attractive
pockets one can observe a series of potential barriers that might be
able to hold three-body resonances.

\subsection{Energies and wave functions.}
\label{subsec2}

As a final step, we now solve the coupled set of differential radial
equations for states with $1/2^+$, $3/2^+$, and $5/2^+$.  Five
adiabatic effective potentials are included in the calculations.
Resonances are then found searching for radial solutions behaving
asymptotically as the outgoing waves specified in Eq.(\ref{eq7}).  To
impose this analytically known asymptotic behaviour is not strictly
necessary.  The complex rotated resonance wave functions are actually
falling off exponentially, and a simple box boundary condition would
then be sufficient to obtain the correct solutions. However this
boundary condition is numerically much more delicate, since the
effective potentials have to be computed accurately at much larger
distances than those required by the analytic boundary condition.

\begin{table}[t]
\vspace*{0.5cm}
\begin{center}
\caption{Resonance energies and widths $(E_{R},\Gamma_{R})$, in MeV, 
for the $1/2^+$, $3/2^+$, and $5/2^+$ states in $^5$H for the Gaussian
three-body effective potential with the range $b_{3b}=3$ fm and three
different strengths $S_{3b}$ given in MeV in the first row.  The last
column shows the results with $S_{3b}=-30$ MeV when only $s$ and
$p$-waves are included.}
\vspace*{0.5cm}
\begin{tabular}{|c|ccc|c|}
\hline
$S_{3b}$         & $-25$ & $-30$   & $-35$   &   $-30 (s,p)$\\
\hline
$1/2^+$  & (1.69, 1.95)& (1.57, 1.53) & (1.40, 1.11)  & (1.55, 1.35) \\
$3/2^+$  & (3.24, 4.31)& (3.25, 3.89) & (3.23, 3.44)  & (3.05, 3.46) \\
$5/2^+$  & (2.85, 3.13)& (2.82, 2.51) & (2.68, 1.83)  & (2.65, 2.00) \\
$5/2^+$  &             &              &               & (3.70, 4.31) \\
\hline
\end{tabular}
\label{tab6}
\end{center}
\end{table}

\begin{table}[t]
\vspace*{0.5cm}
\begin{center}
\caption{Resonance energies and widths $(E_{R},\Gamma_{R})$, in MeV, 
for the $1/2^+$, $3/2^+$, and $5/2^+$ states in $^5$H for the Gaussian
three-body effective potential with strength $S_{3b}=-30$ MeV and four
different ranges $b_{3b}$ given in fm in the first row.  }
\vspace*{0.5cm}
\begin{tabular}{|c|cccc|}
\hline
$b_{3b}$      & $2.95$ & $3.00$   & $3.05$   &  $3.15$  \\
\hline
$1/2^+$  & (1.63, 1.70)& (1.57, 1.53) & (1.51, 1.37)  & (1.36, 1.05) \\
$3/2^+$  & (3.26, 4.09)& (3.25, 3.89) & (3.23, 3.69)  & (3.17, 3.26) \\
$5/2^+$  & (2.86, 2.75)& (2.82, 2.51) & (2.76, 2.26)  & (2.62, 1.79) \\
\hline
\end{tabular}
\label{tab7}
\end{center}
\end{table}

We estimate the range, $b_{3b}$, of the effective Gaussian three-body
potential, $V_{3b}=S_{3b}e^{-\rho^2/b_{3b}^2}$, to be around 3.0 fm
which is the hyperradius corresponding to a configuration where all
the three particles are touching each other.  In table \ref{tab6} we
give the corresponding computed resonances in $^5$H for different
strengths varying around the value $-30$~MeV which places the lowest
resonance energy close to the measured values.  The main effect of
decreasing the three-body attraction is to increase the width of the
resonances, while the energies only increase moderately.  A further
decrease of attraction produce $3/2^+$ and $5/2^+$ resonances too
broad to appear for the scaling angles used in the calculation.

Similar effects are observed when the range of the three-body
potential is changed. The results shown in table \ref{tab7}
correspond to a three-body force with a fixed strength $S_{3b}=-30$
MeV and four different values of the three-body range $b_{3b}$. Again
the main effects are observed in the width of the resonances. A
decrease of $b_{3b}$ gives rise to larger resonance energies and
widths, but a change of 0.2 fm in the range modifies the energies by
at most 0.27 MeV, while the widths change between 0.7 MeV and 1 MeV
for all cases.  Smaller values of $b_{3b}$ produce 3/2$^+$ and 5/2$^+$
resonances too broad to be obtained with the complex scaling angles
used in the calculation.  From tables \ref{tab6} and
\ref{tab7} we conclude that the ground state of $^5$H is a
1/2$^+$ resonance. If the energy is adjusted to be around 1.6 MeV, the
two excited states, 3/2$^+$ and 5/2$^+$, appear about 1.65~MeV and
1.25~MeV higher, respectively.

To test the role played by the $d$-waves, we show in the last column
of table~\ref{tab6} the energies and widths of the resonances when the
two-body $d$-wave potentials vanish.  A strength $S_{3b}=-30$ MeV has
been used for the three-body force in this case.  Inclusion of the
$d$-wave interactions only marginally changes the structures.  The
variation in the energies and widths is certainly visible, and
especially the widths increase by including the $d$-waves.  In fact,
when the $d$-wave interactions are zero a second $5/2^+$ resonance
appears at about 3.70 MeV.  With $d$-waves this resonance is too broad
to be obtained with the scaling angle $\theta=0.30$ rads used in this
calculation.

\begin{figure}
\vspace*{-1.5cm}
\epsfig{file=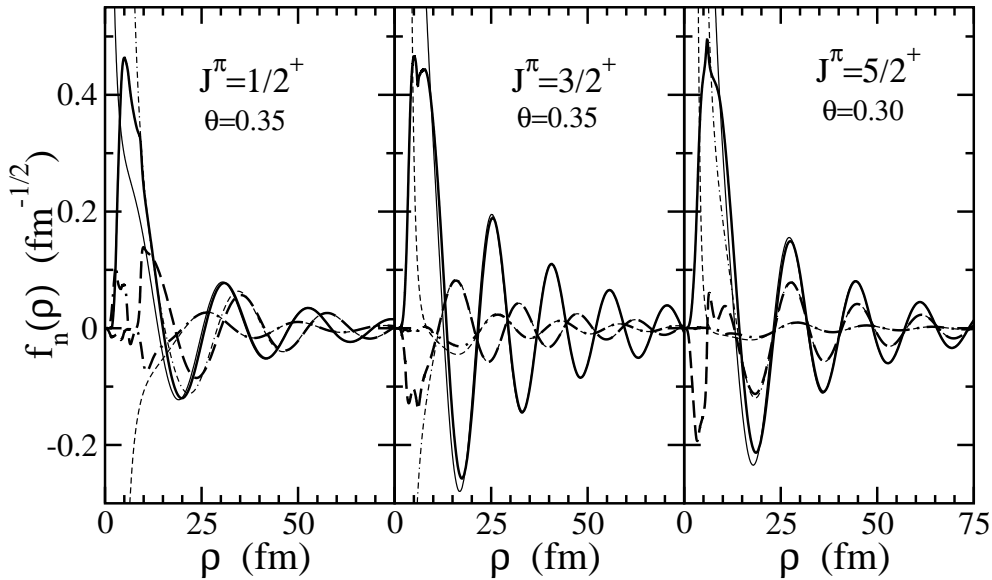, width=11cm, angle=-90}
\vspace*{-1.5cm}
\caption{Real parts of the rotated radial wave functions for the 1/2$^+$ 
state (left), the 3/2$^+$ state (middle) and the 5/2$^+$ state (right)
in $^5$H with a rotation angle $\theta$. We show the computed wave
functions that contribute most (thick lines) and the asymptotic
functions $\sqrt{\rho}H_{K+2}^{(1)}(|\kappa| \rho
e^{i(\theta-\theta_R)})$ (thin lines), see Eq. (\ref{eq7}).}
\label{fig5}
\end{figure}

For the three resonances, the radial wave functions associated with the
three lowest effective potentials given in Fig.\ref{fig4} are shown by
the thick curves in Fig.\ref{fig5}.  The wave functions have been
obtained with a three-body force with Gaussian parameters $S_{3b}=-30$
MeV and $b_{3b}=3.0$ fm. To keep the figure cleaner only the real
parts are shown.  As expected, after complex scaling, the resonance
radial wave functions are approaching zero according to the
asymptotics given in Eq.(\ref{eq7}), which is shown in the figure by
the thin curves. In all cases the computed wave functions have reached
the expected asymptotic behavior already at about 40 fm.

Since the complex rotated three-body wave functions can be normalized,
we compute the relative weights of the different components given in
the last rows of tables \ref{tab3}, \ref{tab4}, and \ref{tab5}. When the
three-body wave functions are written in the first Jacobi set
($\bm{x}$ between the two neutrons) the $1/2^+$ ground state is
clearly dominated by the $\ell_x=\ell_y=L=0$ component (more than 90\%
of the norm), while the $sd$ interferences ($L=2$) are the dominating
components for the excited $3/2^+$ and $5/2^+$ states (75\% and 85\%
of the norm, respectively). In the second and third Jacobi sets
($\bm{x}$ from $^3$H to one of the neutrons) the $\ell_x=\ell_y=1$
components dominate in all the three cases, whereas the coupling to
the total $L$ must produce the same distribution in all Jacobi
systems, i.e. $L=0$ and $L=2$ for ground and excited states,
respectively.  Also the total spin $S$ is conserved in transformations
between Jacobi systems.

\subsection{Comparison with other computed and measured results}

\begin{table}[h]
\vspace*{0.5cm}
\begin{center}
\caption{Theoretical and experimental energies ($E_R$) and widths 
($\Gamma_R$) for the 1/2$^+$, 3/2$^+$ and 5/2$^+$ resonances in
$^5$H. The results correspond to the references quoted in the first
column. The first, second and third rows show our results,
respectively for our full interaction (with $S_{3b} = -30$ MeV $b_{3b}
= 3$ fm), when the $d$-wave interactions are zero, and with the
$^3$H-neutron potential given in
\cite{shu96}. }
\vspace*{0.5cm}
\begin{tabular}{|cccc|}
\hline
\multicolumn{4}{|c|}{$(E_{R},\Gamma_{R})$ (MeV) } \\
\hline
$J^{\pi}$ & 1/2$^+$ & 3/2$^+$ & 5/2$^+$ \\
\hline
$^5$H(full) & (1.57,1.53) & (3.25,3.89) & (2.82,2.51) \\
$^5$H($d=0$) & (1.55,1.35) & (3.05,3.46) & (2.65,2.00) \\
      &             &             & (3.70,4.31) \\
Theor. \cite{shu96} & (2.26,2.93) & (4.41,6.22) & (2.58,1.78) \\
                   &             &             & (3.81,4.70) \\
Theor. \cite{shu00} & (2.5-3.0,3-4) & (6.4-6.9,8) & (4.6-5.0,5) \\
Theor. \cite{des01} & (3.0-3.2,1-4) & (--,--) & (--,--) \\
Theor. \cite{ara03} & (1.59,2.48) & (3.0,4.8) & (2.9,4.1) \\
Exp. \cite{kor01} & ($1.7\pm0.3$,$1.9\pm0.4$) & (--,--) & (--,--) \\
Exp. \cite{sid03} & ($1.8\pm0.1$,$<0.5$) & (--,--) & ($2.7\pm0.1$,$<0.5$) \\
Exp. \cite{gol05} & (1.8,1.3) & ($>2.5$,--) & ($>2.5$,--) \\
Exp. \cite{ter05} & (2,2.5) & ($>2.5$,--) & ($>2.5$,--) \\
Exp. \cite{mei03a} & (3,6) & (--,--) & (--,--) \\
Exp. \cite{gur05} & ($5.5\pm0.2$,$5.4\pm0.6$) & ($>10$,$>2$) & ($>10$,$>2$) \\
\hline
\end{tabular}
\label{tab8}
\end{center}
\end{table}

We compare our results in the first row of table \ref{tab8} to
different theoretical an experimental values extracted from the
literature. The results given in the third row correspond to the
resonances obtained by following the same method as described in this
work, but using the $^3$H-neutron interaction given in
ref.\cite{shu96}. This potential is used in \cite{shu00} to obtain the
$^5$H-resonances (fourth row) after solving the Schr\"{o}dinger
equation by means of a hyperspherical harmonics expansion of the wave
function.  It is not clear whether $d$-wave interactions are included
in the calculations in ref.\cite{shu00}.  For this reason the results
in the second and third rows of table \ref{tab8} omitted $d$-waves.
We then observe that the results given in the third and fourth rows
are distinctly different, even if similar two-body potentials are used
in both cases.  This is due to the fact that in \cite{shu00}
resonances are determined from the observed rapid variations in the
phase shifts as a function of the energy, while we obtain them as
poles of the $S$-matrix. The definition as $S$-matrix poles usually
gives lower energies and smaller widths.  In any case the potential
used in \cite{shu00} is producing a ground state at an energy
significantly higher than with the potential used in this work, and in
fact not very far from the energy of the $5/2^+$ state.  We emphasize
that the potential in \cite{shu96} has the form in Eq.(\ref{eq9}),
that we have found to be inconsistent with the available $^3$H-neutron
and $^3$He-proton experimental data.

In \cite{des01} also a phase shift analysis is employed to get the
three-body resonances, and again the energy obtained for the ground
state is roughly a factor of two higher than obtained in the present
work.  It is remarkable that our energies agree very well with those
of \cite{ara03}, where a microscopic model combined with complex
scaling is used to obtain resonances as poles of the $S$-matrix.
However, our widths are systematically smaller than those of
\cite{ara03}.  This has to be attributed to either different
interactions or contributions from the triton-core.  Among the
different available experimental data, only \cite{gur05} clearly
disagree with our results.  On the other hand, for most of the cases,
the agreement between our theoretical energies and widths and the
experimental values is quite good \cite{kor01,gol05,ter05,sid03}.

\section{Energy distributions}
\label{sec6}

\begin{figure}
\vspace*{-1.5cm}
\epsfig{file=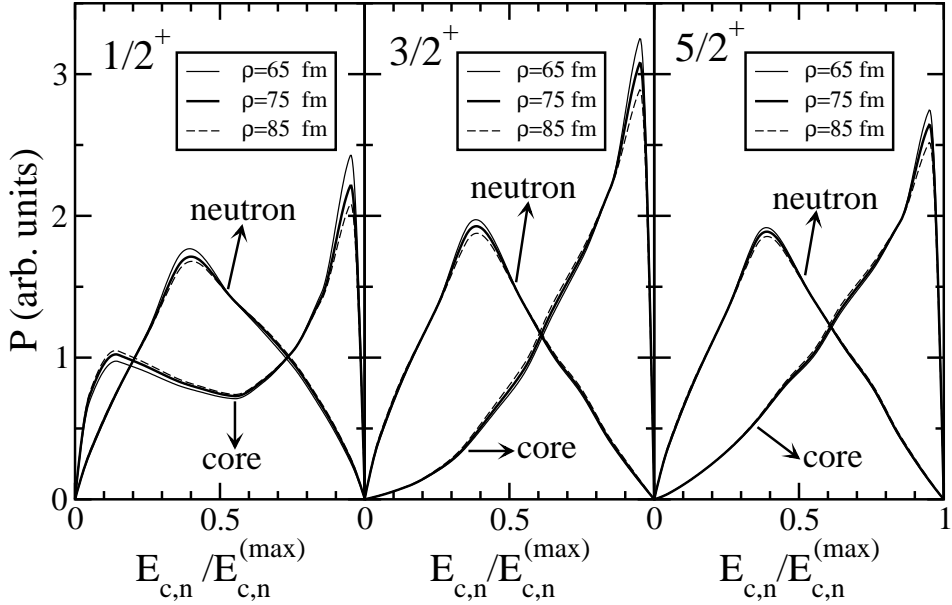, width=11cm, angle=-90}
\vspace*{-1.5cm}
\caption{Energy distributions of the $^3$H-core and the neutrons after decay 
of the 1/2$^+$ (left), 3/2$^+$ (center), and 5/2$^+$ (right)
three-body resonances in $^5$H. The convergence of the results is
illustrated by the different (almost overlapping) curves, which
correspond to calculations with $\rho_{max}$ values of 65 fm (thin
solid), 75 fm (thick solid), and 85 fm (dashed). }
\label{fig6}
\end{figure}

Recently we presented a method to compute the energy distributions 
of particles emerging from a  three-body decay of a many-body resonance 
\cite{fed04,gar06}.  The  method exploits the fact  that hyperspherical 
harmonics transform into themselves after Fourier transformation.
Therefore the kinetic energy distribution of the fragments after decay
of the resonance is, except for a phase-space factor, obtained as the
absolute square of the total wave function in coordinate space for a
large value of the hyperradius, but where the five hyperangles are
interpreted as in momentum space \cite{gar06}.

After integration over the four hyperangles describing the directions
of the two Jacobi momenta, $\bm{k}_x$ and $\bm{k_y}$, conjugate to
$\bm{x}$ and $\bm{y}$, the probability distribution as function of
$k_y^2$ ($k_y^2 \propto \cos^2 \alpha$, where $\alpha$ is the fifth
momentum hyperangle) is given by
\begin{equation}
 P(k_y^2) \propto P(\cos^2\alpha) \propto \sin(2\alpha)
    \int d\Omega_x d\Omega_y|\Psi(\rho_{max},\alpha,\Omega_x,\Omega_y)|^2 \;,
\label{eq11}
\end{equation}
where $\rho_{max}$ refers to a large value of the hyperradius where
the asymptotic behaviour of the three-body wave function has been
reached.  Except for mass factors, $k_y$ is the momentum of the
third particle relative to the three-body center of mass.  Therefore
the kinetic energy of the third particle is proportional to $k_y^2$
(or to $\cos^2 \alpha$), and its energy distribution is then as given
in Eq.(\ref{eq11}). In particular, $\cos^2 \alpha$ gives the energy of
the particle relative to its maximum possible energy in the decay
process.

This procedure has formally the shortcoming that the hyperradius
$\rho_{max}$ in principle should be asymptotically large (since the
detection takes place at large distance), and the larger the value of
$\rho_{max}$ the larger the size of the required basis for the
hyperspherical expansion. For short-range potentials the asymptotic
limit is known to be the hyperspherical harmonics, reflected in the
corresponding hyperharmonic spectrum of the hamiltonian without
interaction.  It was shown in previous reports \cite{gar06} that the
correct asymptotic limit is reached already at intermediate distances
where the basis size is still manageable in the numerical
calculations.  An increase of $\rho_{max}$  requires a larger
basis which then would reproduce the energy distribution found at the
smaller distance with the smaller basis size.  This optimum region of
hyperradii is determined as the region where the observable is
independent of $\rho_{max}$-variations; further increase of
$\rho_{max}$ is not productive.  Clearly this is a satisfactory
procedure for short-range interactions.

Following this procedure we have computed the energy distributions of
the fragments after decay of the $^5$H-resonances. The results for the
1/2$^+$, $3/2^+$, and $5/2^+$ states are shown in the left, central,
and right parts of Fig.\ref{fig6}. Neutron and triton energy
distributions are shown as a function of the corresponding particle
energy relative to its maximum energy. Convergence has been tested by
comparing calculations for three different values of $\rho_{max}$ (65
fm, 75 fm, and 85 fm) in Eq.(\ref{eq11}). As seen in the figure all
three calculations provide almost overlapping curves.

For the 1/2$^+$ resonance (left part of Fig.\ref{fig6}), the neutron
energy distribution has an irregular broad peak at intermediate
energies.  The triton energy distribution has one peak very close to
the maximum energy and a broader peak at low energies.  This pattern
corresponds to two types of decay mechanisms. The first is emission of
$^3$H followed by decay of an intermediate two-neutron structure. This
sequential decay amounts to a two-body process where the $^3$H-core
receives maximum energy and the two neutrons move together in the
opposite direction relative to the core. In the subsequent decay each
neutron then must share the remaining energy which leads to an
intermediate energy between zero and the maximum value. The existence
of a low-lying neutron-neutron virtual $s$-state appears to be
decisive in the decay process, even if a stable intermediate
configuration of two neutrons does not exist, neither as bound states
nor as resonances.

The second decay structure with the small triton energy corresponds to
emission of two neutrons in opposite directions while the triton
essentially remains at rest in the center. The two neutrons then share
the available energy leading to intermediate energies for each of
them.  Combining these two mechanisms produces the computed irregular
broad single-neutron peak.

For the $3/2^+$ and $5/2^+$ states (central and right parts of
Fig.\ref{fig6}) similar distributions are found for the neutrons except
that they tend to be slightly more narrow.  In contrast the triton
energy distribution only has the high energy peak corresponding to the
sequential decay of triton emission.  The low-energy triton peak is
not present because it is unfavorable for two neutrons to be far apart
with a $d$-wave describing their center of mass motion. 

\begin{figure}
\vspace*{-1.5cm}
\epsfig{file=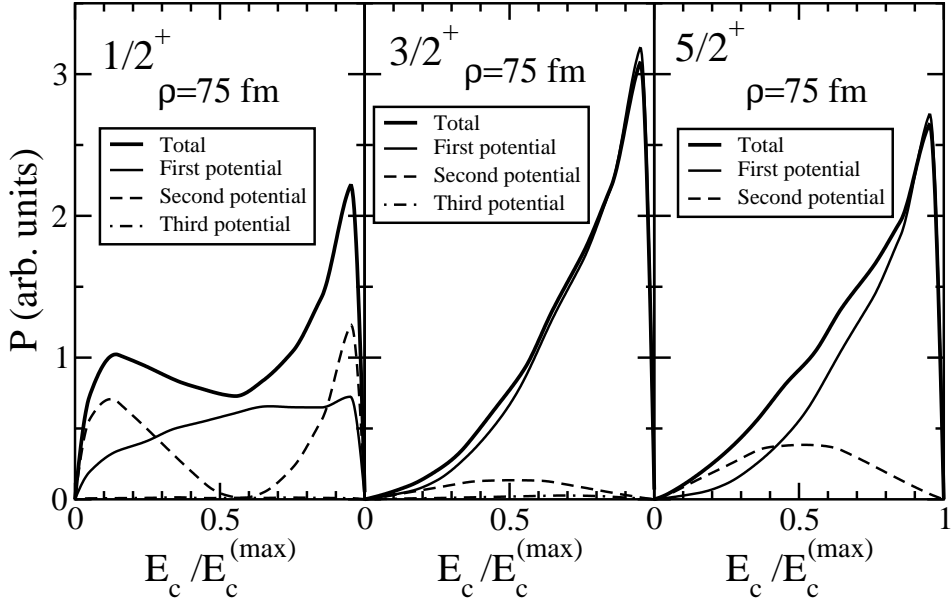, width=11cm, angle=-90}
\vspace*{-1.5cm}
\caption{Contribution from the three adiabatic potentials in Fig.\ref{fig4} to the
triton energy distributions after decay of the 1/2$^+$ (left), 3/2$^+$ (center),
and 5/2$^+$ (right) three-body resonances in $^5$H. The thick solid curve
gives the total distributions, while the thin solid, thin dashed and thin dot-dashed
curves give the individual contributions from the first, second, and third
adiabatic potentials, respectively. In the right part of the figure the contribution 
from the third adiabatic potential is not visible.}
\label{fig7}
\end{figure}

In Fig.\ref{fig7} we show, for the three computed $^5$H resonances, the individual contributions 
to the triton energy distributions from the three adiabatic potentials shown in Fig.\ref{fig4}. 
In the 5/2$^+$ case the contribution from the third potential is not visible.  The results in the figure 
correspond to $\rho_{max}=75$ fm. 
In all the three cases the main features of the total distribution (thick solid 
line) are produced by a single adiabatic potential (the second one for 1/2$^+$, and the first
one for 3/2$^+$ and 5/2$^+$). In particular these potentials are responsible 
for the two peaks at high and small triton energies in the 1/2$^+$ case (left), and for the high energy
peaks in the 3/2$^+$ (center) and 5/2$^+$ cases (right). 

In the three cases, at $\rho= 75$ fm,
the angular eigenfunctions associated to this adiabatic potential is almost 100\% given
by the components in the second columns of tables~\ref{tab3}, \ref{tab4}, and \ref{tab5}, respectively.
These components correspond to a relative $s$-wave between the two neutrons. However, while in the
1/2$^+$ case the triton is also in a relative $s$-wave relative to the two-neutron center of mass,
in the 3/2$^+$ and 5/2$^+$ cases the triton is in a relative $d$-wave. This fact inhibits the appearance
of the low energy peak in the triton energy distribution in the 3/2$^+$ and 5/2$^+$ cases.
For the next most relevant adiabatic potential (the first one for 1/2$^+$, and the second one for 
3/2$^+$ and 5/2$^+$), the contribution comes also from the $\{\ell_x=\ell_y=L=0, S=1/2\}$ component
in the 1/2$^+$ case, but for the 3/2$^+$ and $5/2^+$ states
it comes from the $\{\ell_x=\ell_y=L=1, S=3/2\}$ component ($\bm{x}$ between the neutrons).

\section{Summary and conclusions}
\label{sec7}

The hyperspheric adiabatic expansion method is used to investigate
$^5$H in a three-body model where the $^3$H-core is surrounded by two
neutrons.  Three-body resonances are computed as poles of the
$S$-matrix.  We use the complex scaling method, which provides
resonances as solutions of the Faddeev equations with complex energy
and wave functions falling off exponentially at large distances.

The two-body $^3$H-neutron interaction is built with central,
spin-orbit, and spin-spin terms, where each of the radial form factors
is a sum of two Gaussians. The strengths and ranges of the Gaussians
are adjusted to fit the available $^3$He-proton experimental
data. These data are more reliable and abundant than the $^3$H-neutron
data, and should lead to an appropriate $^3$H-neutron potential simply
by switching off the Coulomb interaction. We have found that the
experimental $^3$He-proton phase shifts can be reproduced only when
the spin operators in the two-body potential are such that the mean
field angular momentum quantum numbers are conserved quantum numbers
for the valence nucleon.  Use of such proper spin operators appears to
be essential in a reliable description.

The ground state of $^5$H is found to have spin and parity $1/2^+$,
and for an energy of around 1.6 MeV the width is about 1.5 MeV. Two
excited states are then found at 2.8 MeV and 3.2 MeV, with spin and
parity $5/2^+$ and $3/2^+$, respectively. Thus these states are broad
and overlapping.  The effective three-body force has a range
corresponding to three touching particles.  Then it modifies mainly
the width of the resonances, keeping the energies rather stable.  The
agreement of these results with most of the available experimental
data is remarkable.

For all the three states the dominating components correspond to a
relative $s$-wave between the two neutrons. In the ground
state this $s$-wave combines with another relative $s$-wave between
the core and the center of mass of the two neutrons, while for the
excited states it combines with a $d$-wave. In the second and third
Jacobi sets the dominating components correspond to relative
$p$-waves between the core and one of the neutrons as well as between
the second neutron and the center of mass of $^4$H.  Finally, we have
found that the $1/2^+$ resonance decays with roughly
equal probabilities through two-body sequential decay by
$^3$H-emission and three-body decay by emission of the two neutrons in
opposite directions.  The $3/2^+$ and $5/2^+$ resonances only decay by
triton emission.  Both neutron and triton energy distributions are
needed simultaneously to interpret these decay modes.

In conclusion, the three-body model describes efficiently the cluster
structure of $^5$H as two neutrons and a triton. The resonances found
are consistent with the experimental data.  To get a good agreement
with the experimental values in both $^5$H and $^4$H, the two-body
neutron-triton interaction must have the spin dependence consistent
with the mean field angular momentum quantum numbers of the triton.
The decay leads to relatively broad energy distributions dominated by
triton emission.

\section*{Acknowledgments}

This work was partly supported by funds provided by DGI of MEC (Spain) 
under contract No. FIS2005-00640.
One of us (R.D.) acknowledges support by a predoctoral I3P fellowship
from CSIC and the European Social Fund.

\end{document}